\documentclass[5p,review,number,british,sort&compress]{elsarticle}
\usepackage{graphicx}
\usepackage{amssymb}
\usepackage{amsmath}
\usepackage{babel}
\usepackage{graphics}
\usepackage{lineno}

\journal{Nuclear Instruments and Methods in Physics Research A}

\begin{document}
\linenumbers
\bibliographystyle{elsarticle-num}
\biboptions{square,comma,sort&compress}

%\bibpunct{[}{]}{,}{n}{,}{,}

\begin{frontmatter}

\title{Muons tomography applied to geosciences and volcanology}

\author[ipnl]{J. Marteau \corref{cor1}}
\ead{marteau@ipnl.in2p3.fr}
\cortext[cor1]{Corresponding author}
%\author[ipnl]{B. Carlus}
%\author[ipnl]{Y. D\'eclais}
\author[ipgp]{D. Gibert} 
\author[ipgp]{N. Lesparre} 
\author[rennes]{F. Nicollin} 
%\author[rennes]{B. Kergosien}
\author[napoli]{P. Noli} 
\author[lhep]{F. Giacoppo} 
                          
\address[ipnl]{Institut de Physique Nucl\'eaire de Lyon (UMR CNRS-IN2P3 5822), Universit\'e Lyon 1, Lyon, France.}
\address[ipgp]{Institut de Physique du Globe de Paris (UMR CNRS 7154), Sorbonne Paris Cit\'e, Paris, France.}
\address[rennes]{G\'eosciences Rennes  (CNRS UMR 6118), Universit\'e Rennes 1, B\^at. 15 Campus de Beaulieu, 35042 Rennes cedex, France.}
\address[napoli]{Universit\`a degli studi di Napoli Federico II \& INFN sez. Napoli, Italy.}
\address[lhep]{Laboratory for High Energy Physics, University of Bern, SidlerStrasse 5, CH-3012 Bern, Switzerland.}

%\thanks{Paper in preparation for \textit{Nuclear Instruments and Methods in Physics Research A}, February 2010.}

\begin{abstract}

Imaging the inner part of large geological targets is an important issue in geosciences with various applications. Different approaches already exist (e.g. gravimetry, electrical tomography) that give access to a wide range of informations but with identified limitations or drawbacks (e.g. intrinsic ambiguity of the inverse problem, time consuming deployment of sensors over large distances). Here we present an alternative and complementary tomography method based on the measurement of the cosmic muons flux attenuation through the geological structures. We detail the basics of this muon tomography with a special emphasis on the photo-active detectors. 
\end{abstract}

\end{frontmatter}

%=========================================================================================================================================
\section{Introduction and motivations}

Monitoring natural events such as earthquakes, volcanic eruptions, landslides and tsunamis has immense importance, both scientific and societal. The interest of volcano radiography arose in the last decades in Japan \cite{nagamine1995geo, nagamine1995method,tanaka2001, tanaka2003}, which has a large volcanic and seismic activity, like other places in the world such as Italy and Iceland in Europe or the Antilles belt in the Atlantic ocean. Because of the possible vicinity of populated areas, volcanoes require careful monitoring of their activity and precise modelling of their geophysical evolution.  
\begin{figure}[!ht] 
   \centering
   \includegraphics[width=7.5cm,height=5cm]{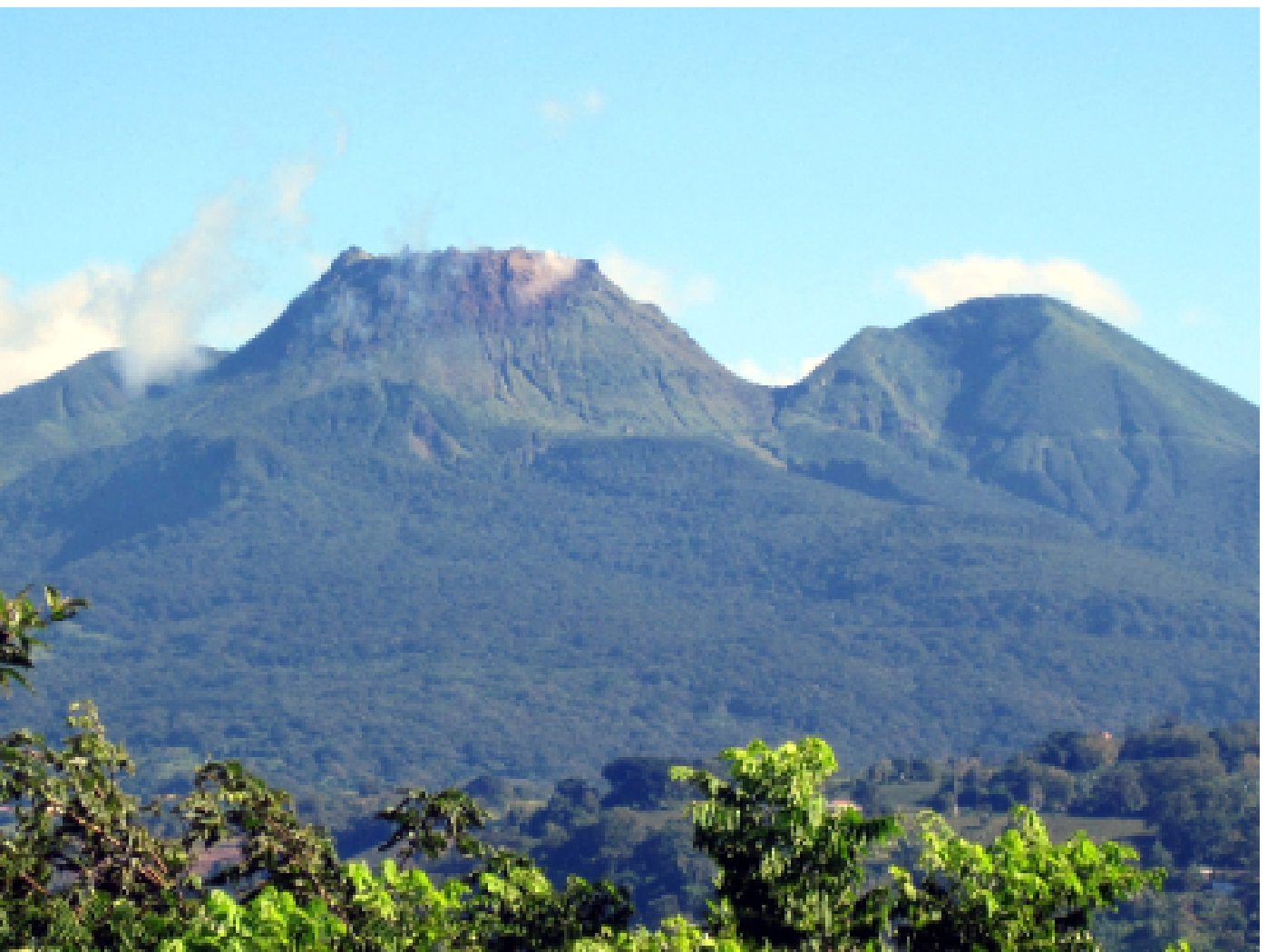} 
   \includegraphics[width=7.5cm,height=5cm]{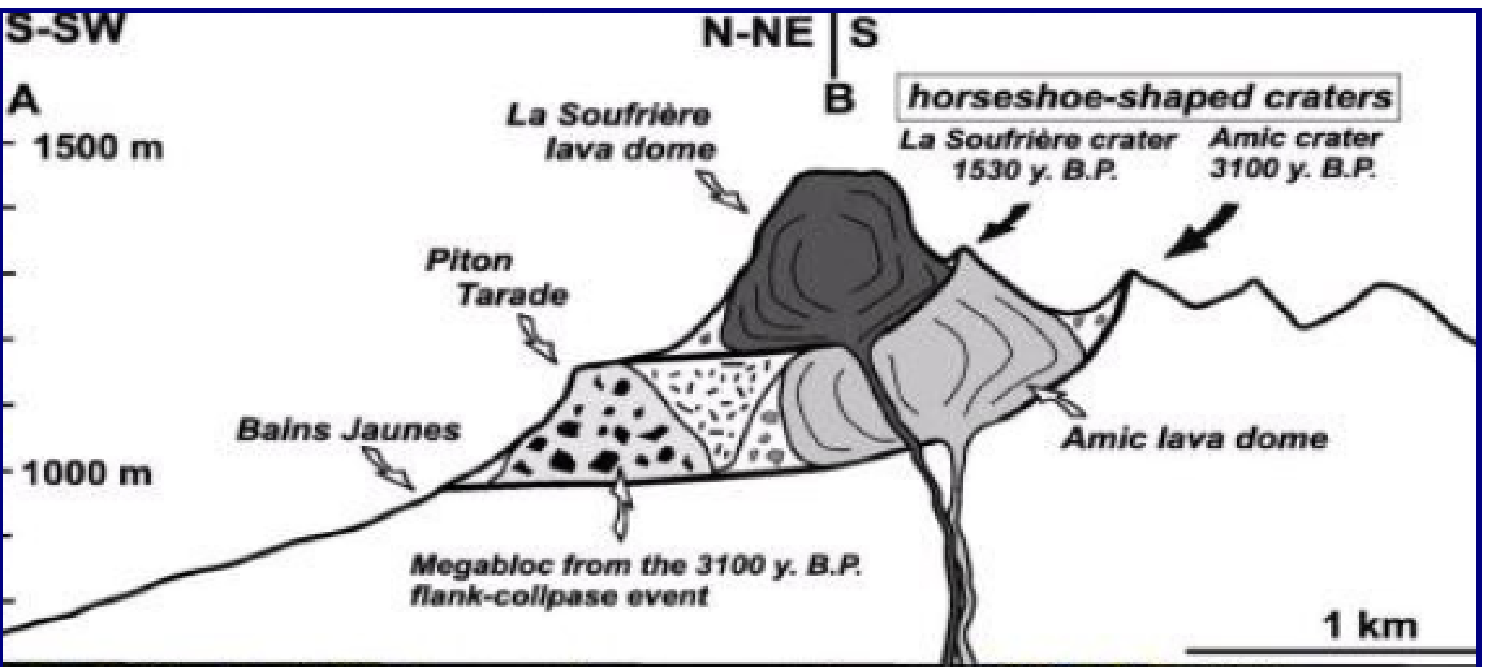} 
%   \caption{Location of Lesser Antilles islands and active volcanoes (top). La Soufri\`ere of Guadeloupe: view from the observatory (middle) and geological model (bottom).}
   \caption{La Soufri\`ere of Guadeloupe: picture and model.}
   \label{Fig:Antilles}
\end{figure}

Consider for instance Lesser Antilles, a subduction volcanic arc with a dozen of active volcanoes located in populated areas. The volcanoes of Martinique (La Montagne Pel\'ee), Guadeloupe (La Soufri\`ere), and Montserrat (The Soufri\`ere Hills) presented an eruptive activity since the beginning of the 20th century. It is therefore crucial to evaluate their eruptive evolution in the near future to and quantify the associated risk for surrounding inhabitants. Reaching these goals requires accurate imaging of the volcano's inner structure and quantitative estimates of the related parameters (variations of volume, density, strain, or pressure) associated with fluid transports (magma, gas, or water) or physical and chemical evolution of the volcanic materials. La Soufri\`ere of Guadeloupe, an andesitic volcano whose lava dome is about five hundred years old \cite{Boudon:2008}, is particularly relevant since it presents a diversified number of hazards including phreatic eruption, flank collapse and explosive magmatic eruption \cite{Komorowski:2008}. Its dome is very heterogeneous, with massive lava volumes embedded in more or less hydrothermalized materials \cite{Nicollin:2006}. Given the constant erosion of the volcano due to the tropical intensive rain activity, the evolution of such a lacunary structure may be rapid, with formation of cavities, that may be filled with pressurized and likely very acid fluids, resulting in flank destabilization. On top of that present structural models show that the dome sits on a 15$^o$ N-S inclined plane, leading to an overall very unstable structure (Fig.\ref{Fig:Antilles}). This particular example shows that a precise knowledge of the dome's internal structure is a key issue for the global modelling and understanding of the volcanoes. For this reason, La Soufri\`ere has been chosen as priority target for muon imaging \cite{Gibert:EPS2010}, which constitutes one of the most promising tools to obtain direct information on the density distribution inside geological objects.

%=========================================================================================================================================
\section{Tomography basics}

The interest of muon tomography for Earth Sciences purposes soon arose after the discovery of cosmic rays and of the muon. The cross-section of that particle at those typical energies makes it a perfect probe since it is able to cross hundredths of meters of rock with an attenuation related to the amount of matter along its trajectory \cite{Nagamine:2003}. Since it is a charged particle, its detection is quite straightforward. The first studies relevant to tomography in geosciences, were motivated by the need to characterise the geological burden overlying underground structures, in particular laboratories hosting large particles experiments aimed at detecting rare events in a silent environment (the so-called ``cosmic silence'' \cite{Zichichi}). This type of ``underground tomography'' is pursued nowadays in the applications of long-term storage where detailed knowledge is required on the geological environment (nature and borders of various layers) and for mining geophysics. Applications other than those directly related to underground physics require smaller, modular, autonomous detectors movable on the field and able to reject efficiently the background. The pioneering archaeological investigations performed in the Egyptian Chephren pyramid %(Fig.\ref{Fig:Alvarez}) 
by Alvarez et al. in the seventies \cite{Alvarez:1970}, looking for some hidden room inside the pyramid, reveal the feasibility of the method.  
%\begin{figure}[ht] 
%   \centering
%   \includegraphics[width=4cm,height=3cm]{figure_alvarez.eps} 
%   \includegraphics[width=4cm,height=3cm]{figure_kefren.eps} 
%   \caption{Alvarez' muon detector (left) and its location within the Chefren pyramid (right).}
%   \label{Fig:Alvarez}
%\end{figure}

\begin{figure}[ht] 
   \centering
   \includegraphics[width=9cm,height=7cm]{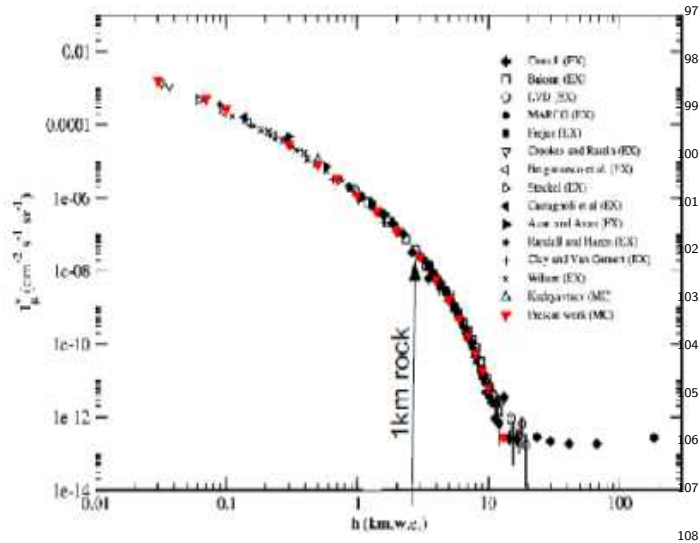} 
   \caption{Integrated flux computed as a function of the standard rock thickness $L$ in meters-water-equivalent (m.w.e.) compared to experimental points.}
   \label{Fig:Flux}
\end{figure}

A muon radiography uses the same basic principles than a standard medical radiography: measuring the attenuation of a beam (cosmic muons versus X-rays) when crossing matter (rock vs human flesh) with a sensitive device. A detailed discussion of all parameters is given in \cite{Gibert:GJI2010}. The measurement gives access to the opacity $\varrho$ of the geological structures by comparing the muons flux $\Phi$ after crossing the target to the incident open sky flux, $\Phi_o$. Various models give analytical expressions of the muon flux from the two-body decays of pions and kaons and assuming a primary proton flux spectrum roughly following a power law $\approx E_p^{-2.7}$ \cite{Bugaev:1970,Bugaev:1998,Gaisser:1990}. The opacity is converted to density $\rho$ by inverting the integral equation~:  \hbox{$\varrho(\mathrm{kg.m}^{-2}) \equiv \int_{L} \rho(\xi) \mathrm{d}\xi,$} $L$ denoting particles trajectory with local coordinate $\xi$. The muons energy loss (and potential absorption) on their way through rock accounts for the standard brem{\ss}trahlung, nuclear interactions, and $e^- e^+$ pair production physical processes, taken as~:
\begin{equation}
-\frac{\mathrm{d}E}{\mathrm{d}\varrho}(\mathrm{MeV}\, \mathrm{g}^{-1} \, \mathrm{cm}^{2}) = a(E) + b(E)E,
\label{EnergyLoss1}
\end{equation}
where the functions $a$ and $b$ depend on the crossed material properties \cite{PDG}. The flux of muons emerging from the target is the integral of $\Phi$ over the energy, ranging from $E_{\mathrm{min}(\varrho)}$, the minimum initial energy necessary to cross given opacity $\varrho$, to infinite (Fig.\ref{Fig:Flux}).
This flux is influenced by various environmental parameters such as altitude, geomagnetic cut-off, solar modulation, atmospheric variations to be accounted for in the simulation models. Finally the number of detected muons is the convolution of the muons flux crossing the target, the data taking duration and the telescope acceptance, which is the key experimental parameter that one may evaluate from the simulation and/or from the data themselves. 

%=========================================================================================================================================
\section{Photo-active detectors for tomography}
\subsection{The DIAPHANE project}

\begin{figure}[ht] 
   \centering
   \includegraphics[width=3.5cm,height=4cm]{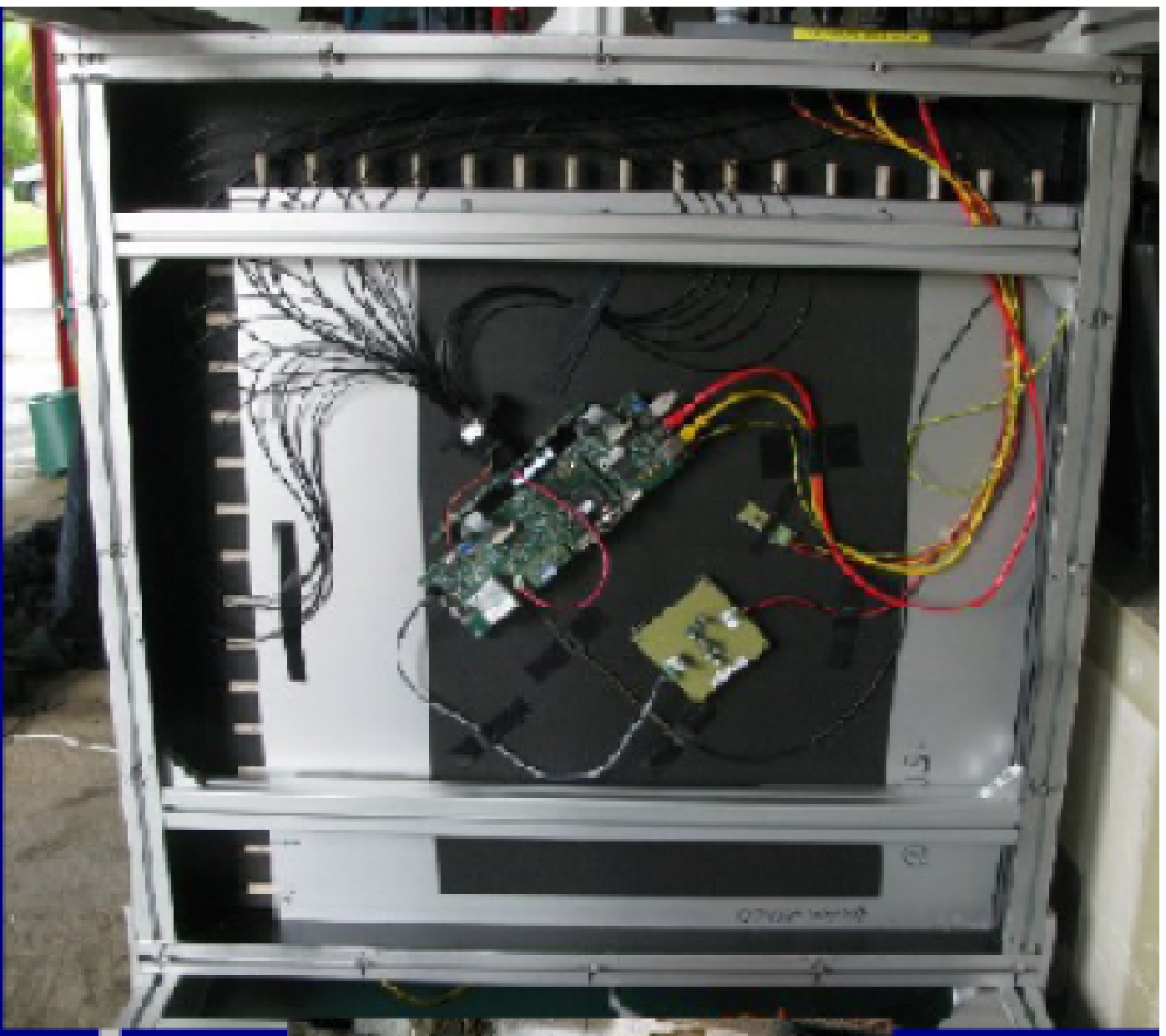} 
   \includegraphics[width=5cm,height=4cm]{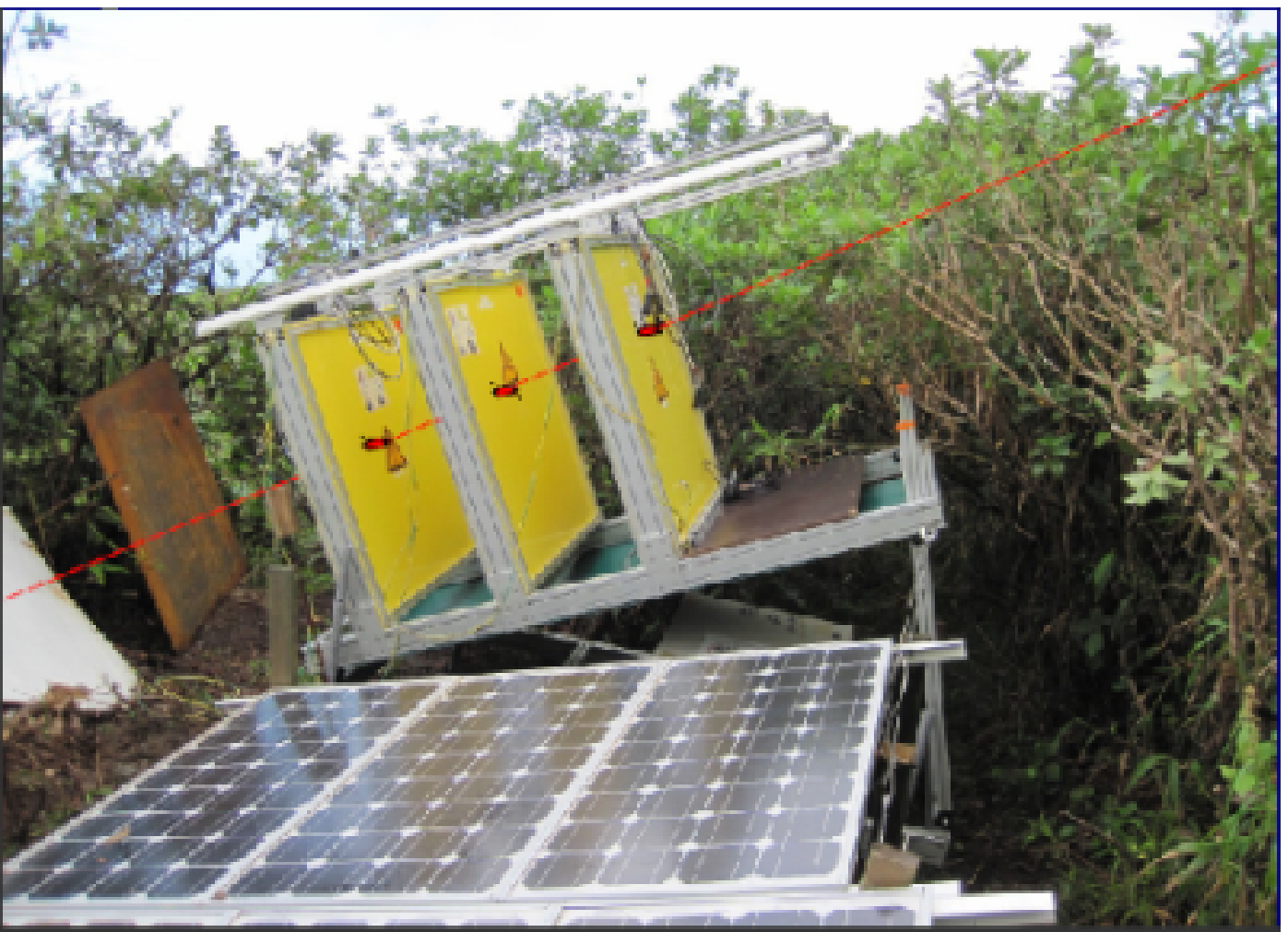} 
   \caption{Left: detection plane with $16\times16$ scintillator bars, connected via optical fibres to the PMT+R/O system. Right: a 3-planes detector installed on La Soufri\`ere (Ravine Sud).}
   \label{Fig:Telescope}
\end{figure}

DIAPHANE is the first european project of tomography applied to volcanology. It started in 2008 with a collaboration between three French institutes~: IPG Paris, IPN Lyon and G\'eosciences Rennes to promote muon tomography in the French Earth Science and Particle Physics communities \cite{Gibert:EPS2010}. The first objectives of the project were to make technological choices for the muon telescopes and to define a design suitable for the difficult field conditions encountered on the Lesser Antilles volcanoes. The detector's design~: plastic scintillator, optical fibres, pixelized photomultipliers and triggerless, smart, Ethernet-capable readout electronics, is based on the state-of-the-art opto-electronics technology, known for its robustness and stability in extreme working conditions. Modularity and limits in weight are also imposed by transportation constraints, some positions on the flank of the volcanoes being accessible only by helicopter. A standard detector (or ``telescope'') comprises 3 independant $XY$ detection planes with autonomous and low power consumption readout system recording and timestamping their own hits in auto-trigger mode. The event-building is performed quasi on-line, via software procedures, by sorting all raw data in time and looking for time coincidences between hits passing the various trigger cuts. Data are transferred continuously via Ethernet wifi and are directly accessible remotely. No shift on-site are needed (concept of the unmanned sensors). The detector is powered through solar panels. Weight, power consumption, robustness and costs have been optimized to the best achievable compromises for that type of field operating detector \cite{Gibert:NIM2010}. 
\paragraph{Detection matrices} Two layers ($X$ \& $Y$) scintillator bars are glued between $1.5 \; \mathrm{mm}$ thick  anodised aluminium plates. The scintillator bars were provided by Fermilab with a rectangular cross-section of $5 \times 1 \; \mathrm{cm}^2$ and are co-extruded with a TiO$_2$ reflective coating and a $1.5 \; \mathrm{mm}$ diameter central hole to host an optical fibre \cite{pladalmau2001}. Two different fibres are used to optimize the emission-absorption spectra matching and decrease the attenuation length~: wavelength shifting (WLS) fibre (Bicron BCF 91A~MC) glued with standard optical cement (Bicron BC-600) in the bar and, through a custom PEEK optical connector, a clear fibre (Bicron BCF-98 MC) down to the photosensor. Three matrices are used in coincidence in a complete telescope. The total aperture angle and the angular resolution of the telescope may be adjusted by changing the distance between the matrices.
\paragraph{Photodetectors} Hamamatsu 64 channels multi-anode photomultipliers are used baseline photosensors (H8804-mod5 and its upgraded version H8804-200mod). These PMTs are robust and do not exhibit any temperature/humidity dependance. Their gains and pedestals are monitored regularly and are stable within a few percents. The present design also foresees optional upgrade with Hamamatsu MPPC (S10362-11-050C) directly connected onto the optical plugs of the scintillator bars w/o the clear fibres. The MPPCs have very attractive performances in terms of single photon sensitivity and photon resolution power, which are key features to improve the muon detection efficiency. Nevertheless their dark count rates and thermal fluctuations are a concern and require careful commissioning. Dedicated electronics, adapted from the PMT's one, is presently under tests. 
\paragraph{Readout system} The global data acquisition system is built as a network of ``smart sensors'' \cite{Girerd:2000,Marteau:2009}. The PMT data are collected by two multichannel front-end chips, then digitized and pre-processed by an Ethernet Controller Module (hosting a 32-bit RISC CPU with a Linux 2.4 OS, a FPGA and a FIFO) plugged on a Controller Mother Board (including a fast ADC, a HV module and a clock decoding system). The same type architecture is also valid for the MPPC option where only the front-end stage has to be adapted. The distributed client/server software is based on the CORBA standard. Since the telescope is running in triggerless mode, event timestamp accuracy is a critical issue. A clock broadcasting system synchronizes all sensors with a common clock unit regulated by GPS. 
\paragraph{Mechanical structure} The frame of the telescope is built with slotted and anodised aluminium profiles. The detection matrices and R/O system fit in a single box made with $4$ profiles and two $1 \; \mathrm{mm}$ thick aluminium plates. Tightness against water and light is obtained with a seal applied between the aluminium plates and the profiles. Four connectors complying with the IP67 norm are used to ensure power supply and data transfer, and a valve equipped with a Gore Tex membrane allows evacuation of water vapour without letting liquid water to penetrate into the box. The supporting structure of the telescope is made with the same type of profiles, the full structure being articulated to change the inclination of the matrices. Pictures of the telescope are shown in Fig.\ref{Fig:Telescope}.

\subsection{First results and comparisons with other methods}    
\begin{figure}[ht] 
   \centering
   \includegraphics[width=4cm,height=4cm]{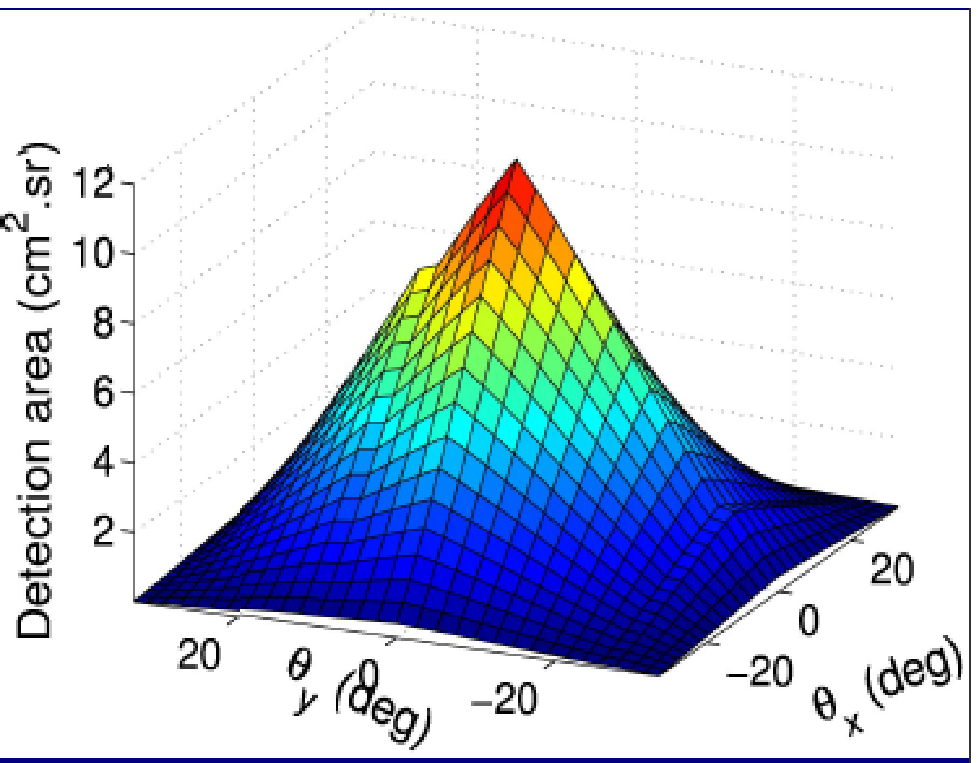} 
   \includegraphics[width=4.5cm,height=4cm]{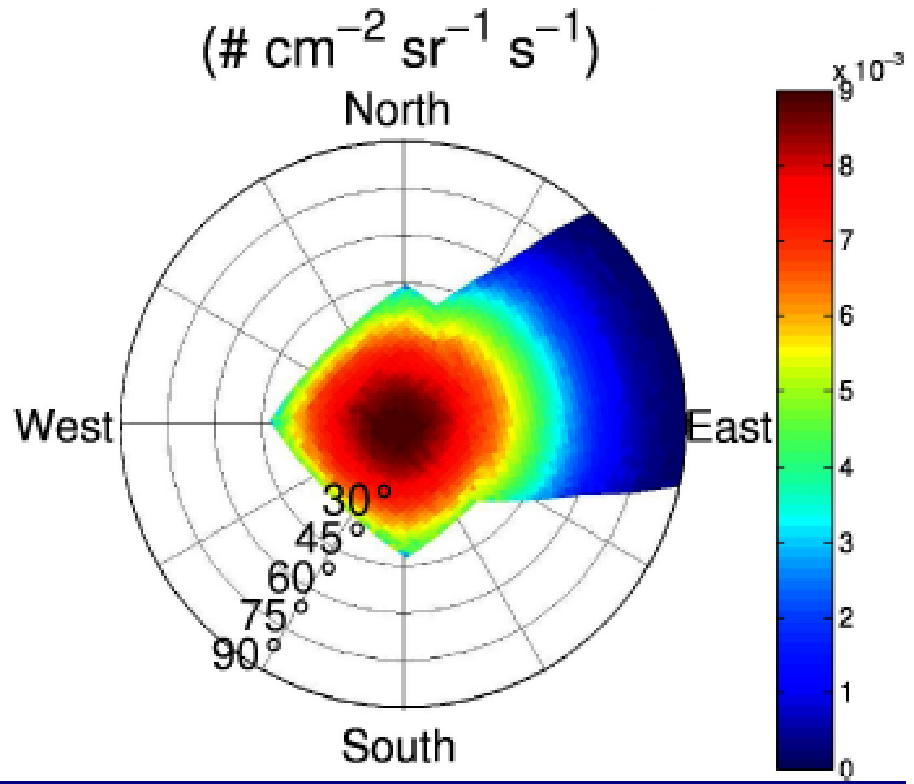} 
   \caption{Left: acceptance function of a $16\times16$ 3-planes detector before correction. Right: corrected open sky muons flux.}
   \label{Fig:Acceptance}
\end{figure}

As stated above, the acceptance of the telescope is a key parameter since the goal of the project is to assign an opacity and therefore a density to the target from an absolute measurement of flux. Many corrections may be inferred to the theoretical acceptance deduced from solid angles calculations. The experimental inefficiencies are corrected either directly from the light yield measurement or indirectly from the overall data sets themselves. Details on the methods may be found in Ref.\cite{Gibert:GJI2010,Gibert:GJI2011}. Typical acceptance curves and corrected open sky muons flux (showing the expected symmetry) are shown in Fig.\ref{Fig:Acceptance}.  

\begin{figure}[ht] 
   \centering
   \includegraphics[width=8cm,height=5cm]{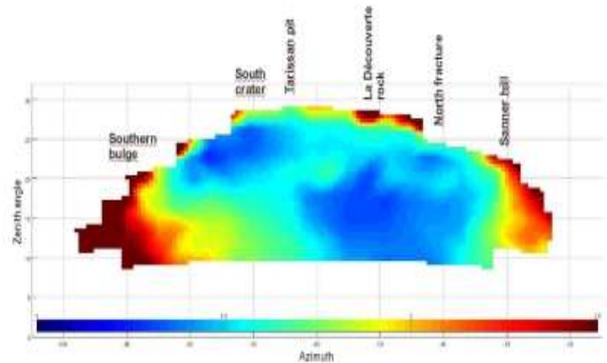} 
   \caption{Density profile obtained at the Roche Fendue site.}
   \label{Fig:TomoSouf1}
\end{figure}

Three DIAPHANE telescopes have been built and have been recording data on the field. The first one was put in the Mont-Terri underground laboratory (Switzerland), located in an anticline formed with layers of Opalinus clay and limestones with densities $\rho_\mathrm{clay} = 2.4$ and $\rho_\mathrm{lime} = 2.7$ \cite{BossartThury:2008}. This place was chosen to fully commission in-situ a muon telescope and constrain detector performance, data analysis and simulations models since the geological layers and topography are well known. This telescope is still taking data and is deployed in various places of the gallery to sample the geological layers and make redundant measurements.\\
The second telescope has been deployed on the Etna volcano for a short trial period during summer 2010 and was able to see within a few days the profile of the volcano. The scale of this volcano implies that, at the current state of the art, only a small portion of its edifice can be investigated through muon imaging. Further campaigns are planed on the Etna following preliminary studies detailed in \cite{Gibert:GJI2010}.\\
The third telescope has been installed on La Soufri\`ere of Guadeloupe, one of the volcanoes with hardest environmental conditions. The geophysics case of this volcano was discussed in the previous sections. Two sites have been already explored, roughly at $90^o$ of each other (``Ravine Sud'', accessible by car and ``Roche Fendue'' only accessible by helicopter). These two orthogonal views show not only a very good compatibility with each other but also with other measurements carried out with different methods on the same place (gravimetry and electrical tomography). Fig.\ref{Fig:TomoSouf1} shows the large density variations observed in the inner structure. Preliminary analysis indicates presence of large low density volumes within the cone, also seen in electrical tomographic data (highly conductive zones being inferred either to hydrothermally washed zones or to acid zones), and reveals the existence of large hydrotherma channels to be accurately monitored.    
The DIAPHANE project has very rich and intense perspectives with the exploration of other fields of view and at a short-term scale the deployment of a network of telescopes (with larger acceptance) running in parallel to perform real-time 3D tomography of the volcano and sample some particularly sensitive zones (like recently opened faults) that may evolve quickly in time.

\subsection{The MURAY project}
The MU-RAY project \cite{Muray:Web} aims at the construction of muon telescopes with angular resolution comparable with that obtainable with emulsions, but with real-time data acquisition and larger sensitive area. The telescopes are required to be able to work in harsh environment imposing a modular structure, each module being light enough to be easily transported by hand. Further requirements are mechanical robustness and easy installation. Power budget must fit a small solar panel system's capability. Good time resolution can improve background suppression by measuring the muon time of flight. 
\begin{figure}[ht] 
   \centering
   \includegraphics[width=2.5cm,height=2cm]{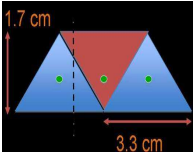} 
   \includegraphics[width=5cm,height=3.5cm]{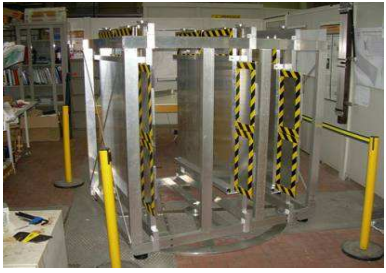} 
   \caption{Left: triangular shape of the scintillator bars used in the MURAY project. Right: a 3-planes detector in the lab.}
   \label{Fig:Muray}
\end{figure}
\paragraph{Scintillator features} A telescope prototype is built in Naples University laboratory and it consists of three $1\times 1 \mathrm{m}^2$$XY$ stations (Fig.\ref{Fig:Muray}). The third station will be used to study possible backgrounds, as the one induced by cosmic-ray showers. Each station is made by two planes disposed in orthogonal way. Each plane is composed by two adjacent modules made by 32 triangular plastic scintillator strips. Scintillator bars are produced by extrusion process at Fermilab, for D0 \cite{Baringer:2001} and Minerva \cite{Minerva:Web} experiments, in pieces as long as 6m, with a hole along the center. The core (Dow Styron 663W) is doped with blue-emitting fluorescent compounds (PPO 1\% and POPOP 0.03\%). The surface has a co-extruded TiO2 coating (0.25mm thick) to increase internal reflectivity and to shield from environment light. The use of isosceles triangular shape allows the construction of very compact, crack-free planes. Moreover, the measurement of the light output produced by two adjacent strips enables the determination of the particle crossing distance between two contiguous fibers, improving the spatial resolution. The triangular bars are glued to each other and over two 2mm thick fibreglass plates, creating a very solid module. Light from scintillator is collected by 1mm diameter WLS fibre BICRON BCF92, glued inside the bar to maximize light collection efficiency. Fibres are mirrored at one end using the Al sputtering facility of the Frascati INFN laboratory \cite{Frascati:Web}.
\paragraph{Photodetectors and electronics } The light is readout by silicon photomultipliers (SiPM) \cite{Sadygov:1998,Golovin:1999} which offer several advantages. Their robustness is mandatory for the environmental conditions; their very low power consumption (less than 1mW per channel) is relevant due to limited power budget. One of main SiPM drawbacks is the gain temperature dependence that affects the detector performance. For this reason the SiPMs' temperature will be controlled using Peltier cells. In order to optimize the power consumption, we decided to group together 32 die SiPMs in a single connector (PCB). The fibres are glued to a custom 32-channel optical connector, which will be fixed to the module chassis and mechanically coupled with the PCB. One side of the Peltier cells is thermally in contact with the back side of the PCB while plastic guarantees a good thermal insulation with respect to the environmental temperature. A rubber O-ring around the sensitive area is used to ensure light and air tightness. Two temperature sensors are located on the PCB for the Peltier cells control circuit. The SiPM front-end electronics readout is based on  SiPM Read-Out Chip (SPIROC) ASIC developed by OMEGA group (LAL, CNRS-IN2P3 \cite{Omega:Web}).
\paragraph{Geophysics case} Today around 600,000 people are living at the base and along the slopes of the Vesuvius volcano, in a  so-called "red" area which has been classified at the highest volcanic risk in Europe. Mt. Vesuvius is therefore among the most studied volcanoes in the world. The knowledge of the inner structure of the volcano edifice and subsoil structure is of the greatest importance to build realistic scenarios of the next eruption through accurate simulations (magma upraising mechanism and eruption). Even if the Vesuvius has been thoroughly investigated using the traditional geophysical methods (gravimetric,  seismological,  electromagnetic), muon radiography may help by improving the resolution of the cone inner structure by one order of magnitude. 

%=========================================================================================================================================
\section{Other techniques : gaseous detectors and nuclear emulsions}
\subsection{Gaseous detectors}
Since there is not an unique way to detect muons, various projects arose recently using gaseous detectors: glass RPC for the TOMUVOL project (radiography of the Puy-de-D\^ome, Clermont-Ferrand, France) and gas TPC for the T2DM2 project (hydrogeology of the karst complex around the LSBB, Rustrel, France). Those projects have taken their first data this year and interesting results are awaited soon.

\subsection{Probing matter with emulsions}
Nuclear emulsion particle detectors feature incomparable high spatial and angular resolution ($<$ 1 ${\mu}m$ and a few $mrad$, respectively) in the measurement of ionizing particles tracks. With the advent of fast electronic detectors, the emulsion technique has experienced a period of decline up to 20 years ago since when impressive progress in the high-speed automated scanning and industrial production have determined a new boost in the application of this technique in high-energy physics experiments. A nuclear emulsion is essentially a photographic plate where silver halide crystals with a typical size of 0.2 ${\mu}m$ are homogeneously dispersed in a gelatin matrix of about 50 ${\mu}m$ thickness. When such an emulsion is exposed to ionizing radiation or light, clusters of silver atoms are produced. These form latent image centres that became visible under an optical microscope when they are reduced to metallic silver filaments (grains) through a chemical developing process. A typical emulsion film produced industrially by FUJI Film Co., as a result of a joint R\&D program with the OPERA Collaboration \cite{Nakamura:06}, consists of $\sim10^{14}$ silver halide crystals. Each of them has a detection efficiency of 20\% for minimum ionizing particle and a sensitivity of $30\div40$ halide grains per 100 ${\mu}m$.

\begin{figure}[ht] 
   \centering
   \includegraphics[width=6cm,height=3.5cm]{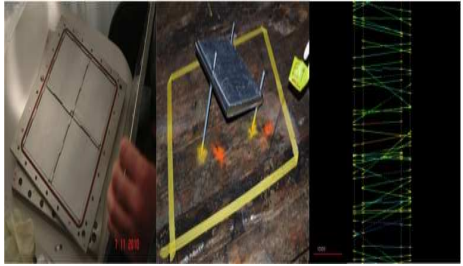} 
   \caption{Left: emulsion module hosting 16 emulsions. Middle: module station during data taking. Right: reconstructed passing-through tracks.}
   \label{fig:bern}
\end{figure}

In addition to their spatial and angular resolution, tracker detectors based on nuclear emulsions are ideal for muon radiography for their data storage capability, portability and rather simple implementation in difficult environments such as, for examples, volcanoes. Moreover, nuclear emulsion films do not need power supply and electronic front-end readout systems.
\begin{figure}[ht] 
   \centering
   \includegraphics[width=8cm,height=5cm]{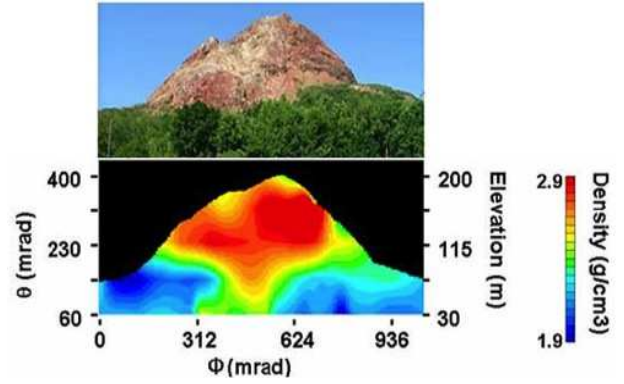} 
   \caption{Top: view of the Showa-Shinzan lava dome. Bottom: average density distribution projected onto the detector's plane.}
   \label{fig:japan}
\end{figure}
The high spatial resolution of nuclear emulsion films was first exploited by Tanaka and his co-workers for the muon radiography of some volcanoes in Japan. In 2007 they performed a test measurement for imaging the conduit of the Showa-Shinzan lava dome, on the east flank of Usu volcano by using quasi-horizontal cosmic-ray muons \cite{Tanaka:07}. A muon detector consisting in a set of emulsion chambers with an area of 6000 $cm^2$ was exposed for three months. Fig.~\ref{fig:japan} shows the reconstructed average density of the dome summit. Recently the Laboratory for High Energy Physics (LHEP) of the University of Bern has started an R\&D program on nuclear emulsion detectors for muon radiography in the framework of the Innovative Nuclear Emulsion Technologies (INET) project, financed by the Switzerland-Russian Scientific and Technological Cooperation Programme. A proof-of-principle test has been conceived in 2010 aimed at the detection of an existing (and known) mineral deposit inside a mine. Dedicated modular devices have been designed (see Fig.~\ref{fig:bern}) and ten samples (emulsion film total area of 5000 $cm^2$) have been placed along the mine tunnel in order to measure the underground muon flux at different locations. The combination of data from several modules would eventually lead to a 3D image of the inner structure of the mountain.

Each emulsion chamber consists of two rectangular stainless steel covers, containing four stacks of two emulsion doublets each. In order to reduce the effects of radiation coming from natural radioactivity, a lead plate is positioned in the middle of each stack. The module mechanics is conceived to be tight to preserve the emulsion surface from light and humidity. After an exposure of 4.5 months the detector have been disassembled \textit{in situ} inside a portable dark room and the emulsion films developed and sent back to Bern. 
LHEP is one of the largest emulsion scanning laboratories in the world where 5 state-of-the-art high speed automatic microscopes are installed and routinely operating for the OPERA experiment  \cite{Acquafredda:09}. In addition, a new scanning station has been dedicated to muon radiography. The automated scanning system consists of a microscope equipped with a computer-controlled motorized stage, a dedicated optical system and a CMOS camera \cite{Arrabito:08}. For each field of view, several tomographic images of the emulsions are taken at equidistant depths by moving the focal plane across the emulsion thickness ($Z$ direction). Images are grabbed and processed by a vision multi-processor board, hosted in the control PC. The tracks are then reconstructed by combining grains from different layers with a dedicated software program (see Fig.~\ref{fig:bern}-(right)). For the extension of muon radiography measurements to deeper structure as well as for reducing the exposure time, a larger emulsion surface would be needed. For this reason an upgrade of the current scanning system is foreseen, in order to reduce the scanning time. One promising solution seems to be the use of high speed GPU (Graphic Processing Unit) to replace standard CPU for faster track reconstruction. In the framework of INET, the LHEP group has set up a facility for ``in-house'' pouring and development of emulsion films. Several tests are ongoing with the aim of producing suitable emulsion detectors (large area, high sensitivity, low noise) for muon radiography and other applications, especially in the medical field. Connections with specialized companies providing emulsion gel have been established for potential future large-scale production.

%=========================================================================================================================================
\section{Conclusions and perspectives}
Muon tomography reaches a new era with mature, robust, adapted to harsh field conditions technologies developed and commissioned around the world. Volcanoes are more and more targeted by muon tomographic imaging since the alternative methods often reveal limited or difficult to work out in their environments. The complementarity of the muon tomography with gravimetric or electric tomography is emerging strongly since it offers direct volume information with quasi straight lines of responses relatively easy to invert. The measurements are also taken in a rather short timescale with a limited number of ``shootings''. Promising real-time 3D tomographies may offer to the community a powerful tool to monitor, understand and better predict the behaviour of volcanoes, with an obvious and crucial societal impact. 

%=========================================================================================================================================
\section*{Acknowledgements}
Authors from the DIAPHANE project warmly thank K.Mahiouz, F.Mounier, P.Rolland, S.Vanzetto (opto-mechanics), B.Carlus (informatics), S.Gardien, C.Girerd, B.Kergosien (electronics) and colleagues from the Observatoire Volcanologique et Sismologique de Guadeloupe~: A.Bosson, F.Randriamora, T.Kitou, C.Lambert and V.Daniel. The DIAPHANE project is financially supported by the IPGP BQR grant, the DOMOSCAN ANR project, the CNRS/IN2P3 Astroparticles program, and the MD experiment of the Mont Terri project funded by Swisstopo and CRIEPI partners. The author would like to thank C.Carloganu and P.Salin who contributed to the oral presentation. Emulsions tomography project benefits from the strong support and expertise of the LHEP group.

\end{document}